\documentclass[letterpaper, 10 pt, conference]{ieeeconf}  
\IEEEoverridecommandlockouts                                                        \usepackage{epsfig} % for postscript graphics files
\usepackage{graphicx}
\usepackage{amsmath}
\usepackage{array}
\usepackage{amssymb}
\usepackage{verbatim}
\usepackage{cleveref}
\usepackage{color}
\usepackage{xcolor}
\usepackage{relsize}
\usepackage{tikz}
\usepackage{fancyhdr}
\usepackage{graphics}
\usepackage{setspace}
\usepackage{epstopdf}
\usepackage{multirow}
\usepackage{lipsum}
\usepackage{textcomp}
\usepackage{mathtools}
\usepackage{cite}
\usepackage{setspace}
\usepackage{booktabs}
\usepackage[noend]{algpseudocode}
\usepackage{flushend}
\usepackage{multicol}
\usepackage{float}
\usepackage{mathtools,bm}
\usepackage[noend]{algpseudocode}
\usepackage{booktabs}
\usepackage{dblfloatfix}
\usepackage{mathrsfs}
\usepackage{mathtools}% Loads amsmath
\usepackage{optidef}
\usepackage{soul}
\usepackage[ruled,vlined]{algorithm2e}

\fancypagestyle{firstpage}{\lhead{\vspace{-0.75 cm} \small \textbf{{\fontfamily{cmss}\selectfont
2020 American control Conference (ACC)\\July 1-–3, 2020, Denver, CO, USA}}}}

\title{\Large \bf
Control Co-design of a Hydrokinetic Turbine: \\A Comparative Study of Open-loop Optimal Control and Feedback Control}

\author{Mohammad Reza Amini$^{1}$, Boxi Jiang$^{1}$, Yingqian Liao$^{2}$, Kartik Naik$^{1}$, Joaquim R.R.A. Martins$^{2}$,  and Jing Sun$^{1}$  
\thanks{$^{1}$M.R. Amini, B. Jiang, K. Naik, and J. Sun are with the Department of Naval Architecture \& Marine Engineering, University of Michigan, Ann Arbor, MI 48109 USA. 
Emails: {\tt\small \{mamini,boxij,naikk,jingsun\}@umich.edu}}
\thanks{$^{2}$Y. Liao and J.R.R.A. Martins are with  the Department of Aerospace Engineering, University of Michigan, Ann Arbor, MI 48109 USA. Emails: {\tt\small \{yqliao,jrram\}@umich.edu}}%
}

\renewcommand{\baselinestretch}{0.95}
\begin{document}

\maketitle
\thispagestyle{empty}
%\thispagestyle{firstpage}%<------------------

%\doublespacing
%%%%%%%%%%%%%%%%%%%%%%%%
\begin{abstract}
Control co-design (CCD) explores physical and control design spaces simultaneously to optimize a system's performance. A commonly used CCD framework aims to achieve open-loop optimal control (OLOC) trajectory while optimizing the physical design variables subject to constraints on control and design parameters. In this study, in contrast with the conventional CCD methods based on OLOC schemes, we present a CCD formulation that explicitly considers a feedback controller. In the formulation, we consider two control laws based on proportional linear and quadratic state feedback, where the control gain is optimized. The simulation results show that the OLOC trajectory could be approximated by a feedback controller. While the total energy generated from the CCD with a feedback controller is slightly lower than that of the CCD with OLOC, it results in a much simpler control structure and more robust performance in the presence of uncertainties and disturbances, making it suitable for real-time control. The study in this paper investigates the performance of optimal hydrokinetic turbine design with a feedback controller in the presence of uncertainties and disturbances to demonstrate the benefits and highlight challenges associated with incorporating the feedback controller explicitly in the CCD stage.
\end{abstract}

%%%%%%%%%%%%%%%
%\vspace{-0.1cm}
\section{INTRODUCTION}
Most complex engineering systems are inherently multidisciplinary in nature. More often than not, the subsystems associated with each discipline exhibit \emph{coupling} with the optimal values of one set of decision variables depending on the selected values of another set. Past research has shown that using conventional sequential optimization methods for such systems may lead to system-level sub-optimality or constraint violation \cite{fathy2001coupling,peters2009measures,patil2012computationally}. Control co-design (CCD) aims to co-optimize the plant and controller of a system while accounting for the coupling between the two \cite{garcia2019control}.

Control co-design frameworks have been widely used to study and optimize the physical design and control system of renewable energy systems in the past, see \cite{ross2022development,kimball2022results,deese2018nested,deodhar2016framework,baheri2017combined,naik2021fused,jiang2022control,pao2021control}, and the references therein.~
In this work, we focus on hydrokinetic turbines (HKT), where the plant optimization aims to maximize the turbine blade's hydrodynamic performance while the control optimization aims to maximize energy generation. The objective function of the optimization problem directly depends on both the blade geometry (defined by the plant decision variables) and the rotational speed (modulated by the control decision variable). In our previous work \cite{jiang2022control}, we identified the control constraint as a strong coupling between the optimal plant design variables and optimal control trajectory, which motivates a CCD-based solution strategy.~The HKT CCD results in \cite{jiang2022control} demonstrated that in the presence of plant-controller design space coupling, CCD leads to a different final design as compared to a sequential design optimization approach, with higher energy generation. Similar to most existing CCD formulations \cite{pao2021control, deshmukh2013simultaneous, naik2022fused, sundarrajan2021open, du2020control, coe2020initial}, the design optimization approach in \cite{jiang2022control} considers CCD with open-loop optimal control (OLOC), where the control trajectory is co-optimized with design space variables.

While CCD frameworks with OLOC schemes provide insights into studying and tackling the intrinsic plant-controller design space coupling, such control schemes are not realizable. Alternatively, \cite{naik2022combined} presents a method to incorporates closed-loop performance into the CCD framework using {proxy functions}. Such a method is however computationally expensive (if not prohibitive) for a large set of plant variables. Using feedback control in CCD explicitly facilitates a design process with a realizable control scheme while also adding inherent robustness to disturbances and uncertainties, as shown in \cite{deshmukh2015bridging, nash2021robust}. Furthermore, previous works in CCD with feedback control have shown minor degradation in performance when compared to OLOC schemes under no disturbances or uncertainties \cite{deshmukh2015bridging, nash2021robust}. However, how to identify the control architecture and parameterize the closed-loop control for CCD remain open problems in general.

In this work, we compare the OLOC schemes with two feedback control architectures (linear and quadratic state feedback) and show that the closed-loop control architectures have comparable system performance (within 1$\%$). Furthermore, we investigate the robustness of the presented control architectures to uncertainties in the system, where the feedback control schemes unsurprisingly outperform the OLOC schemes. To solidify the previous statement and demonstrate the efficacy of the proposed feedback control architecture, we compare the control performances of the closed-loop formulation and an OLOC with knowledge of the uncertainty.

The main contributions of this work are twofold. Firstly, we introduce two realizable feedback control architectures and the associated CCD formulations. Building on our previous work \cite{jiang2022control}, we demonstrate the efficacy of the proposed feedback control schemes using the previously developed OLOC scheme as the performance benchmark. Second, we perform a comparative study of the control performances of each controller under various flow uncertainties to show that selected feedback control scheme performs comparably with an omniscient controller (OLOC with knowledge of uncertainty). 

The rest of the paper is organized as follows: Sec. \ref{sec:sec_2} summarizes the HKT system, which includes the plant and control system. Sec. \ref{sec:sec_3} presents the CCD optimization formulation for HKT. Sec. \ref{sec:sec_4} presents the CCD results and compares the performances of the presented formulations. Sec. \ref{sec:sec_5} presents a sensitivity analysis to flow uncertainties to assess the robustness of the presented control schemes. Finally, the conclusions are summarized in Sec. \ref{sec:sec_6}. 

\vspace{-3pt}
\section{Rotor Blade Design and Control for Maximum Power Production}\label{sec:sec_2}
\begin{figure}[h!]
\centering
      \includegraphics[scale=0.5]{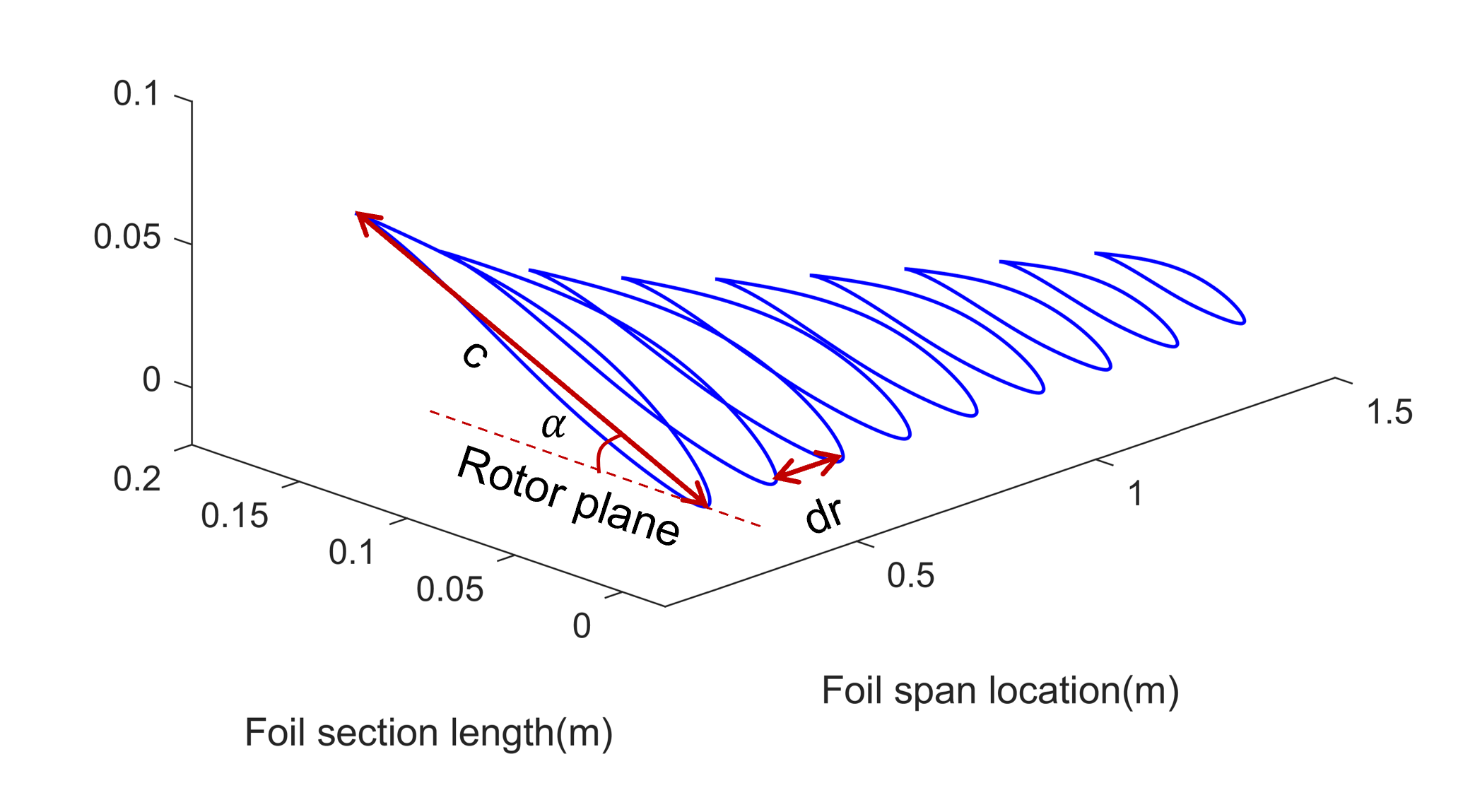}\vspace{-6pt}
      \caption{The notional representation of blade geometry and model in BEM. {$c$: chord length, $\alpha$: twist angle, and $dr$: length of discretized elements along radius.}}\vspace{-12pt}
            \label{figure: geom}
\end{figure}

\vspace{-2pt}
The main objective for HKT system design and control spaces optimization is to maximize its energy production in a time-varying water flow condition. For the baseline design, a free-stream three-bladed horizontal-axis HKT geometry is adopted from~\cite{bahaj2007power}, which is briefly introduced in Sec.~\ref{sec:sec1_A}. To simulate and predict rotor performances with specific geometric design variables, a quasi-static model based on Blade Element Momentum (BEM)~\cite{ning2014simple} theory is used in this study. As shown in Fig.~\ref{figure: geom}, BEM divides a blade into several sections and each section is referred to as a blade element. The details of the design optimization problem formulation are discussed in Sec.~\ref{sec:sec_3}.

\vspace{-5pt}

\vspace{-6pt}
\subsection{Baseline HKT Model}\label{sec:sec1_A}
The baseline HKT geometry is adopted from~\cite{bahaj2007power}. A geometric scaling ratio of 3.5:1 is applied so that the scaled model has a rated power of 5 kW, selected for remote riverine applications by the ARPA-E SHARKS program~\cite{garcia2021engineering}. The geometric and kinematic properties are maintained to be the same between the original Bahaj model~\cite{bahaj2007power} and the scaled model. The rotor performance is assumed to be Reynolds invariant. The detailed parameters of the original Bahaj model and the scaled HKT model used in this study are listed in Table~\ref{table_1}.
\vspace{-4pt}
\begin{table}[h!]
\small
\caption{Specifications of the Bahaj turbine and the scaled HKT model used in this study.} \vspace{-10pt}
\label{table_1}
\begin{center}
\begin{tabular}{l c c}
\hline
 & \textbf{Bahaj} & \textbf{Scaled}\\
 & \textbf{Turbine~\cite{bahaj2007power}}& \textbf{HKT Model} \\
\hline
Rated Power ($kW$) & 0.4  & 5\\
Rated Speed ($m/s$) & 1.5 & 1.7\\
Rotor diameter ($m$) & 0.8 & 2.8\\
Number of blades (-) &3 &3\\
\hline
\end{tabular}\vspace{-10pt}
\end{center}
\end{table}

\vspace{-3pt}
\subsection{Rotor Design Variables}
The twist angles and chord lengths along the blade are selected in this paper as the geometry variables for HKT design space optimization. Other design variables, e.g., thickness to chord length ratio along the radius and blade radius, are kept constant. {The turbine is simplified to have a single foil section, the NACA63815 profile that is used in the Bahaj turbine~\cite{bahaj2007power}. To define the design variables of turbine rotors, the blades are first discretized into nine segments. The turbine mass and rotational inertia are calculated with the assumption that the turbine blades are solid and made with aluminum 6061.}

\vspace{-3pt}
\subsection{CCD with Open-loop Optimal Control (OLOC)}\label{sec:OLPC}
A variable-speed control is implemented to track the turbine's maximum power coefficient locus when the HKT is operating in the below-rated region~\cite{kim2012maximum}. The turbine maximum power coefficient itself is defined by the physical design of HKT. We first consider an open-loop optimal control (OLOC) for torque regulation~\cite{kim2010fast}, assuming that the generator torque can be controlled directly for optimal power production. In the initial design optimization stage, we also assume that the accurate inflow velocity profile is known; this assumption will be relaxed in Sec.~\ref{sec:sec_5} when studying the system performance under uncertainties in inflow conditions. The continuous OLOC trajectory is discretized in time using the third-order Legendre-Gauss-Lobatto collocation method and then solved numerically as a nonlinear programming problem in Dymos~\cite{falck2021dymos}. Generator torque at each time instance is treated as the control space optimization variable ($u$). Different scenarios with and without a constraint on the generator torque are further analyzed in Sec.~\ref{sec:sec_3}.
%%%%%%%%%%%%%%%%%%%%%%%%%%%%%%%%%%%%%%%%%%%%%%%%%%%%%%%%%%%%%%%%%%%%%%

\vspace{-3pt}
\subsection{CCD with Feedback Controller}
In addition to OLOC formulation, a new CCD formation is considered, where a feedback controller is implemented for generator torque control. Both HKT rotational speed ($\omega$) and the square of the rotational speed ($\omega^2$) are considered for feedback control. We note that the previous studies have demonstrated that the ``$K\omega^2$'' control law enables the turbine to approach optimal tip speed ratio tracking at below rated region~\cite{Abbas2022,Bossanyi2000,Johnson2006}. In our CCD formulation, the feedback gain is the control space design variable. The main difference between CCD with OLOC and CCD with a feedback controller is that in the former the control trajectory is optimized, while in the latter a constant gain is optimized to optimize the mechanical power output. The latter formulation reduces the number of control design variables from $N^{*}$ (the number of discrete-time sampling points) to 1. The feedback controllers with the linear rotational speed ($u_l$) and the square of the rotational speed ($u_q$) are defined as follows, respectively:
\vspace{-4pt}
\begin{gather}\label{eq:FB_controller}
    u_l = K_1 \omega, \\
    u_q = K_2 \omega^2\label{eq:FB_controller_s},
\end{gather}
where $K_1$ and $K_2$ are fixed feedback gains that will be optimized during the CCD process. Similar to Sec.~\ref{sec:OLPC} with OLOC, different scenarios with and without a limit on the generator torque are also analyzed in Sec.~\ref{sec:sec_3}. 

Note that the CCD with feedback controllers in~(\ref{eq:FB_controller}) and (\ref{eq:FB_controller_s}) cannot handle constraints on $u$ explicitly and saturation function may need to be employed. {To better condition the CCD optimization problem with constraints, and avoid numerical difficulties that could be resulting from the discontinuity in derivatives}, instead of the standard saturation function ($sat()$):
\vspace{-4pt}
\begin{gather}\label{eq:sat}
    sat(u,\gamma) = \begin{cases} u,~~~~~~~~~~~~|u| \leq \gamma \\ \gamma sign(u),~~~|u|\geq\gamma, \end{cases}
\end{gather}
a smoothed version of the saturation function ($\tilde{sat}()$) is adopted and modified from~\cite{avvakumov2000boundary} and used in the paper to limit the control signal $u\in[0, \gamma]$ as follows:
\vspace{-2pt}
\begin{gather}\label{eq:sat_sat}
    \tilde{sat}(u,\gamma,\mu) = \frac{\gamma}{4}\left(2 + \sqrt{\nu+(\frac{2u}{\gamma})^2}-\sqrt{\nu+(\frac{2u}{\gamma}-2)^2}\right),
\end{gather}
where $\gamma$ is the limit on $u$, and $\nu$ is a small positive number. With $\nu$ set to be 0.001, the smoothed saturation function in (\ref{eq:sat_sat}) is able to approximate the original saturation function in (\ref{eq:sat}) with an error of $<1\%$ of the limit.

\vspace{-4pt}
\subsection{Optimizer}
We used SNOPT~\cite{gill2005snopt}, a well-known sequential quadratic programming algorithm, as the optimizer with both feasibility tolerance and optimally tolerance of $10^{-8}$ for CCD optimization.

\vspace{-4pt}
\section{Control Co-design}\label{sec:sec_3}

A coupled model consisting of hydrodynamic analysis and optimal torque control is used to perform CCD optimization. The CCD conducts design optimization by simultaneously co-optimizing twist angle $\alpha$ and chord length $c$ along the radius, as well as the control variable, $u$, which is defined to be the generator torque. The power $P$ generated by the rotor is given by 
\vspace{-4pt}
\begin{gather}
    P = Q \omega,
\end{gather}
where the flow-induced torque $Q$ is a function of velocity \textit{v}, $c_i,~\alpha_i$, and tip speed ratio $TSR=\frac{\omega r}{v}$, where \textit{r} is the turbine radius. For HKT rotational speed dynamics ($\dot{\omega}$), a simple dynamic model is used which assumes the drivetrain is rigid and its energy loss is negligible:
\vspace{-4pt}
\begin{gather}
    \dot{\omega}=\frac{Q-u}{I_{turbine}},
\end{gather}
where $I_{turbine}$ is the rotor inertia that depends on chords. In this study, the generator dynamics are ignored. 

\begin{figure}[h!]
\centering
      \includegraphics[scale=0.35]{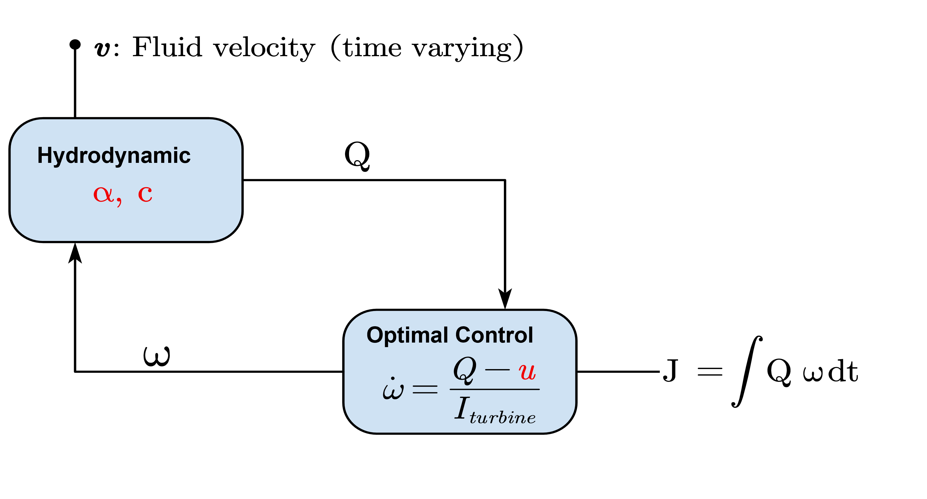}\vspace{-20pt}
      \caption{Schematic of the CCD optimization process.}\vspace{-8pt}
       \label{fig: CCD}
\end{figure}

The schematic of the CCD with OLOC process for the time period $[0,t]$ is shown in Fig.~\ref{fig: CCD}, where the CCD formulation is given by 
\vspace{-4pt}
\begin{maxi}|l|
    {u_{[0,t]},c_i,\alpha_i}{J= \int_{0} ^t Q \omega dt}{}{}
    \addConstraint{\dot{\omega}}{=\frac{Q-u}{I_{turbine}}}
    \addConstraint{Q}{=g(c_i,\alpha_i, v ,TSR)}
    \addConstraint{0.01~m}{< c_i \leq 1~m}{\quad i = 1,2,..., N}
    \addConstraint{-30^\circ}{\leq \alpha_i \leq 30^\circ}{\quad i = 1,2,..., N}
    \addConstraint{0}{\leq u_{[0,t]} \leq u_{max}}
    \addConstraint{0}{\leq \omega },
    \label{Equ:CCD}
\end{maxi}
where, the function $g$ captures the fluid-induced torque based on blade element momentum theory using CCBlade~\cite{ning2013ccblade}. $N$ is the number of discretized sections on turbine blades and $u_{max}$ is the upper limit on generator torque. The turbine is constrained to have a positive rotating speed, i.e., it rotates in one direction only. 
\vspace{-12pt}

For the CCD case with a feedback controller, the optimization formulation in (\ref{Equ:CCD}) is slightly updated as follows:
\vspace{-4pt}
\begin{maxi}|l|
    {K_j,c_i,\alpha_i}{J= \int_{0} ^t Q \omega dt}{}{}
    \addConstraint{\dot{\omega}}{=\frac{Q-u}{I_{turbine}}}
    \addConstraint{u}{=\tilde{sat}(K_j \omega^j,u_{max},\mu)}
    \addConstraint{Q}{=g(c_i,\alpha_i, v ,TSR)}
    \addConstraint{0.01~m}{< c_i \leq 1~m}{\quad i = 1,2,..., N}
    \addConstraint{-30^\circ}{\leq \alpha_i \leq 30^\circ}{\quad i = 1,2,..., N}
    %\addConstraint{0}{\leq u_{[0,t]} \leq u_{max}}
    \addConstraint{0}{\leq \omega },
    \label{Equ:CCD_FB}
\end{maxi}
where $j=1$ for the controller with the linear $\omega$ as feedback (\ref{eq:FB_controller}), and $j=2$ for the case with the square of $\omega$ (\ref{eq:FB_controller_s}) as the feedback, i.e., quadratic feedback controller.

\vspace{-4pt}
\section{Comparative Case Studies}\vspace{-3pt}
\label{sec:sec_4}
We showed in our previous study \cite{jiang2022control} that, for the case study defined in Sec.~\ref{sec:sec_3} and in the absence of a constraint on the control input, the sequential and CCD design optimizations would lead to the same final rotor geometry, revealing the insights on the coupling between the control and physical design spaces. We consider two scenarios in this section. In Scenario 1, we perform the CCD with OLOC and with feedback control.~ In scenario 2, we perform the same design optimization as in Scenario 1 while imposing the control load constraint.

\subsection{Scenario 1: No Constraint on Control Load}\label{sec:sce1}
\begin{figure}[h!]
\centering
      \includegraphics[scale=0.6]{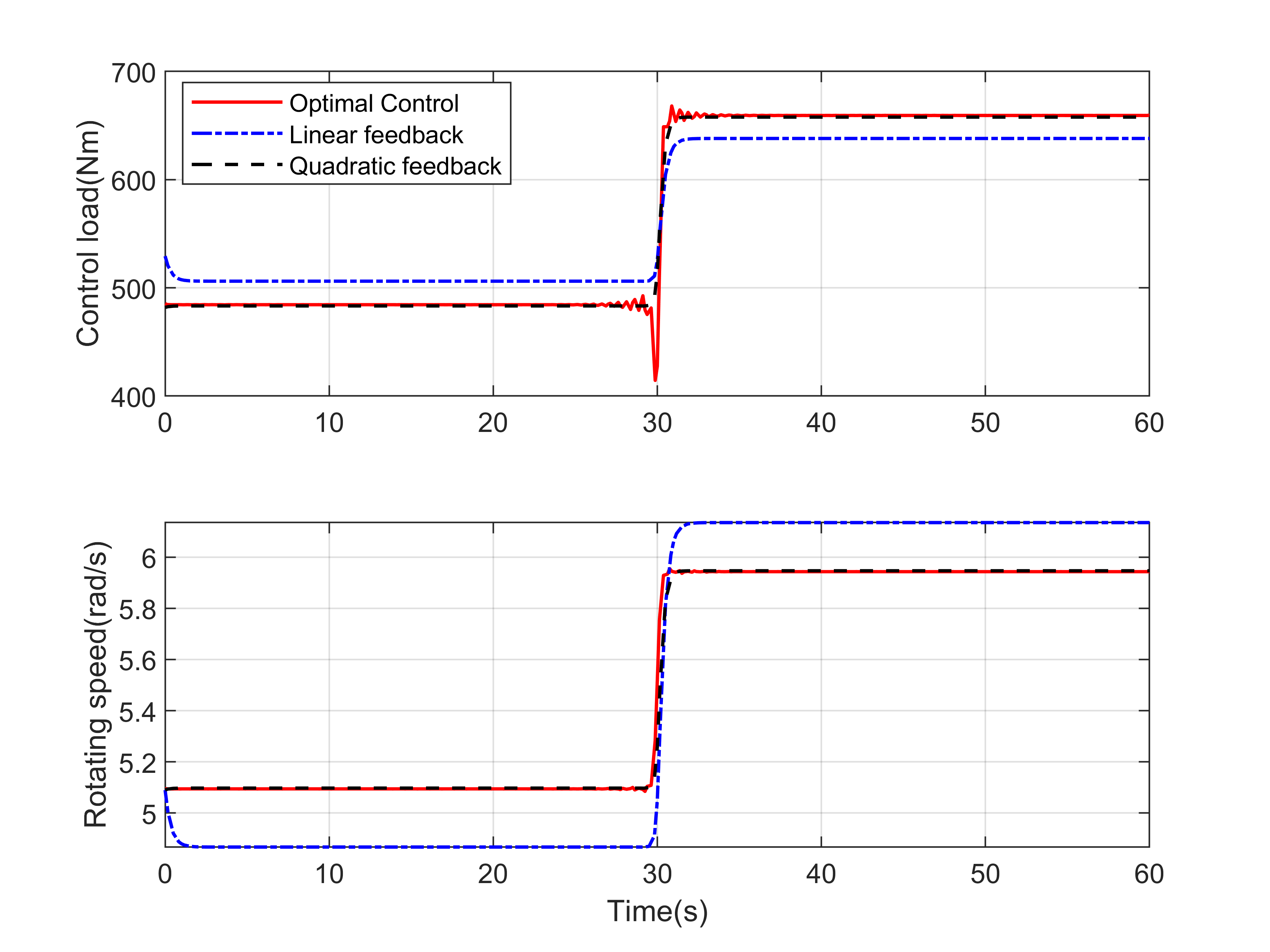}\vspace{-15pt}
      \caption{Control load (top) and HKT rotating speed (bottom) for CCD with OLOC, linear feedback, and quadratic feedback, optimized for inflow condition defined in (\ref{eq:simple_flow}) .}\vspace{-15pt}
       \label{fig:Fig3}
\end{figure}

First, let's consider a simple hypothetical inflow condition which is defined as
\begin{gather}\label{eq:simple_flow}
    v(t) = 1.2 + \frac{0.2}{1+e^{(-(t-30))}}.
\end{gather}
The inflow condition defined in (\ref{eq:simple_flow}) represents a smoothed step change in a flow with an initial speed of 1.2 $m/s$ that changes to 1.4 $m/s$ at around $t=30~s$. Fig.~\ref{fig:Fig3} shows the control load and the turbine rotational speed {trajectories} for CCD with OLOC, linear feedback (\ref{eq:FB_controller}), and quadratic feedback (\ref{eq:FB_controller_s}). It can be seen that, unlike the linear feedback controller, the quadratic feedback controller has similar steady-state performance (where the rotational speed is constant before and after the step change) compared to the case with OLOC. The optimized linear and quadratic controller gains ($K_1$, $K_2$) are listed in Table~\ref{table_2}.
\vspace{-8pt}

\begin{table}[htbp!]
\small
\caption{Feedback controller gains and total energy output of CCD without control load constraint for inflow condition (\ref{eq:simple_flow}).} \vspace{-10pt}
\label{table_2}
\begin{center}
\begin{tabular}{l c c}
\hline
\textbf{CCD Approach} & \textbf{Control Gain} & \textbf{Energy Output}
\\
 & $K_1,~K_2$ & [$J$] \\
%  & $[Nm/red/s]2$ & \\
\hline
OLOC & N/A  & 191563 (\textbf{ref})\\\hline
Linear & 104.03 & 191378 (-0.1\%)\\
Feedback Control& {$[Nm/(rad/s)]$} & \\\hline
Quadratic & 10.98 & 191560 (-0.002\%)\\
Feedback Control& {$[Nm/(rad/s)^2]$} & \\
\hline
\end{tabular}\vspace{-10pt}
\end{center}
\end{table}

As listed in Table~\ref{table_2}, the CCD with OLOC has the highest energy output. The designs with linear and quadratic feedback controllers generate 0.1\% and 0.002\% less energy, respectively, than the OLOC case. The results shown in Fig.~\ref{fig:Fig3} and Table~\ref{table_2} indicate that for the hypothetical inflow condition (\ref{eq:simple_flow}), the OLOC trajectory can be effectively approximated with the CCD framework by both linear and quadratic feedback controllers, with the quadratic feedback controller showing the best approximation results.

\begin{figure}[h!]
\centering
      \includegraphics[scale=0.5]{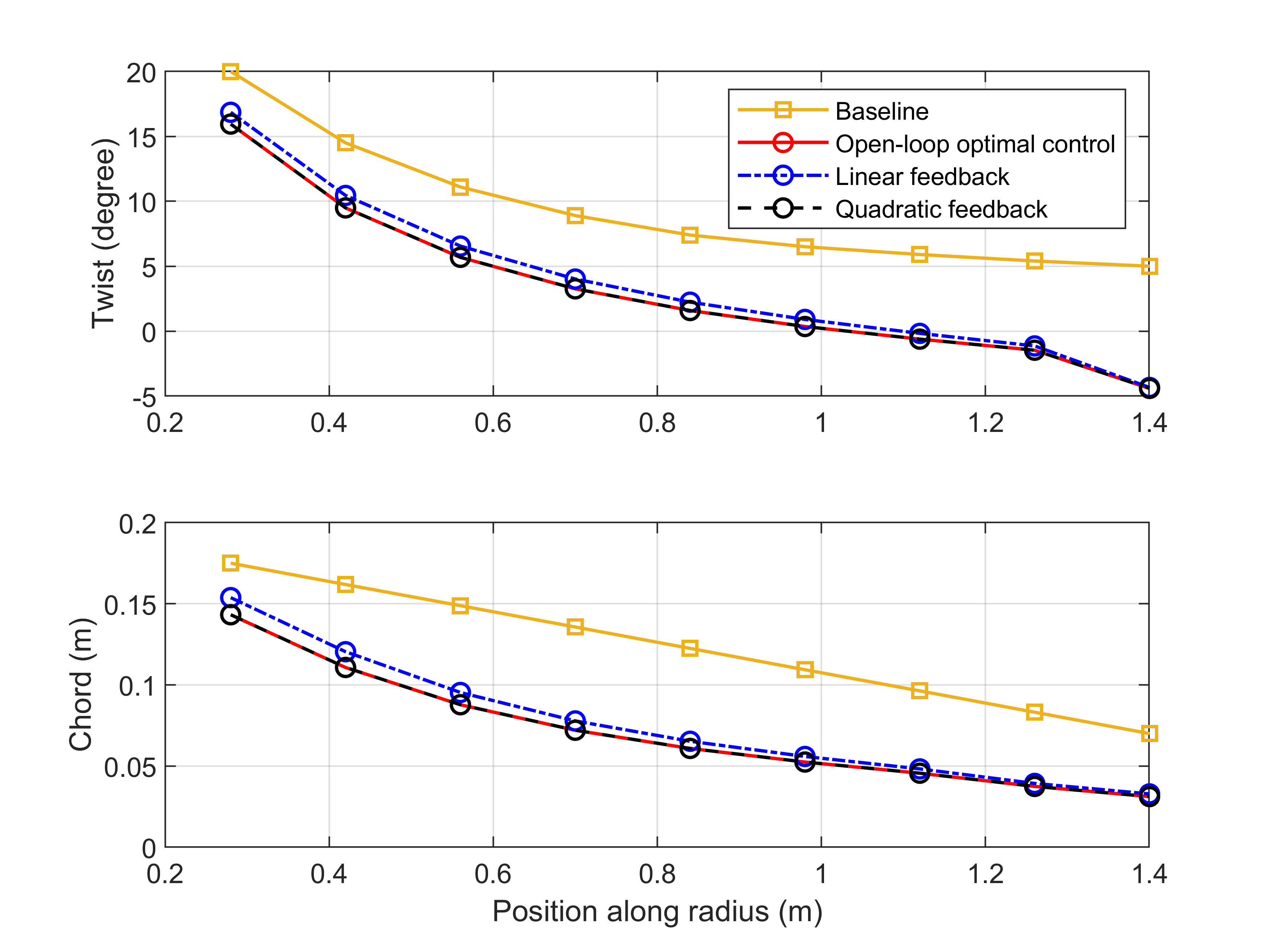}\vspace{-12pt}
      \caption{Comparison of physical geometry optimized by CCD  with OLOC, and linear and quadratic feedback controllers, as well as the baseline design without constraint on control for inflow condition defined in (\ref{eq:flow_sin_base}). (top) twist angle, and (bottom) chord length.}\vspace{-10pt}
       \label{fig:Fig4}
\end{figure}

For a more complex inflow condition defined as follows
\vspace{-4pt}
\begin{gather}\label{eq:flow_sin_base}
    v_0(t) = 0.2 \sin(0.25t) + 1.5. 
\end{gather}
The CCD process with OLOC and linear and quadratic feedback controllers are repeated. The physical geometry results, the twist angle and the chord length distributions, are plotted in Fig.~\ref{fig:Fig4} and compared against the baseline design. Fig.~\ref{fig:Fig5} also shows the HKT rotating speed, tip speed ratio, control load, and fluid-induced torque resulting from CCD optimization with the three controllers. The optimal gains of the feedback controllers, as well as the total energy output from CCD with these three controllers are listed in Table~\ref{table_3}. 
%
%\vspace{-6pt}

\begin{figure}[h!]
\centering
      \includegraphics[scale=0.5]{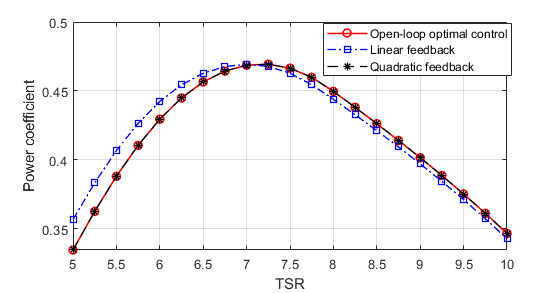}\vspace{-10pt}
      \caption{{Power coefficients versus TSR for CCD optimization with OLOC, linear and quadratic feedback with no constraint on control load.}}\vspace{-10pt}
       \label{fig:cp1}
\end{figure}
%
%\vspace{-6pt}
\begin{figure}[h!]
\centering
      \includegraphics[scale=0.6]{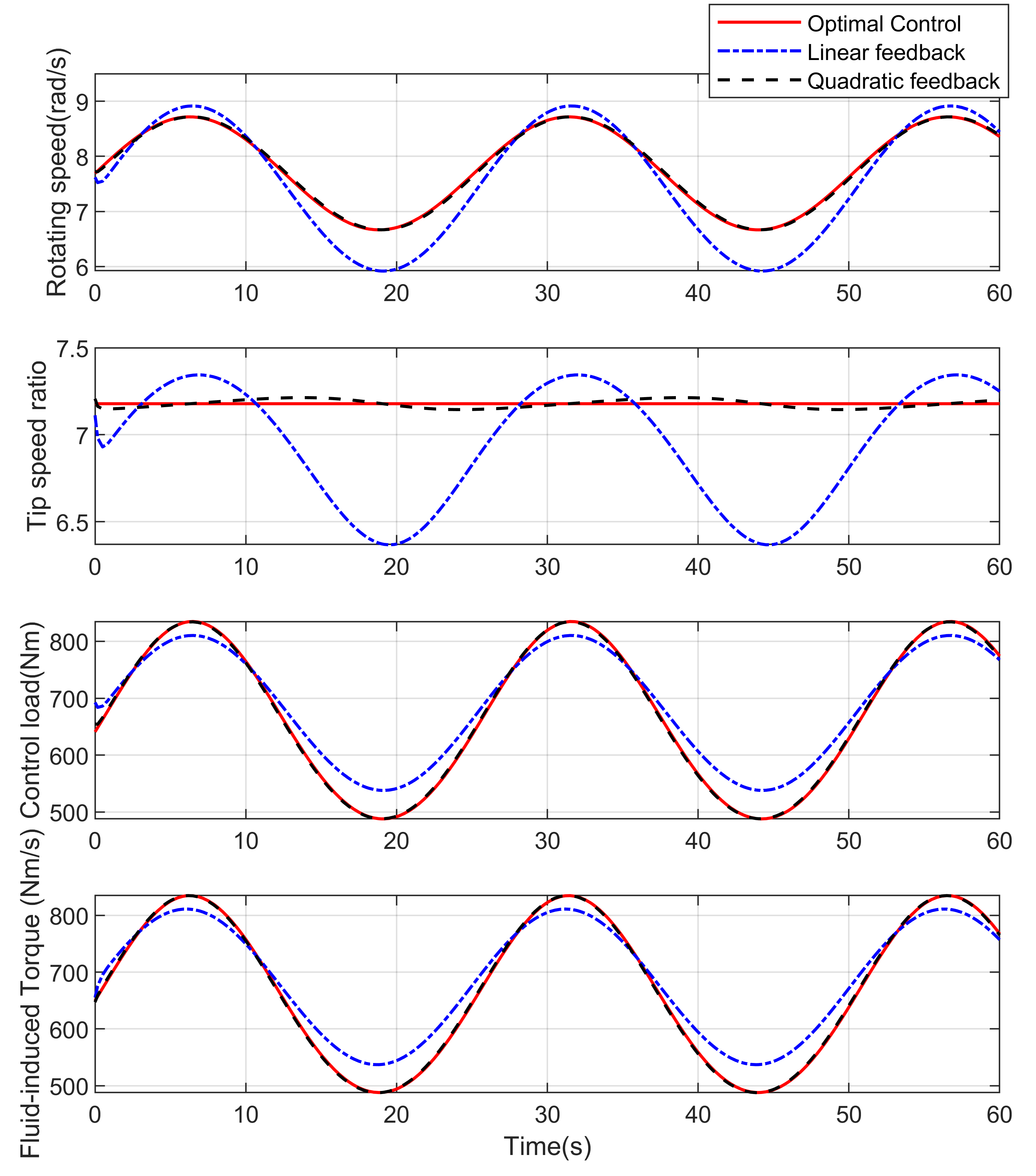}\vspace{-5pt}
      \caption{HKT rotating speed, tip speed ratio, control load, and fluid-induced torque resulted from CCD optimization with OLOC, linear feedback, and quadratic feedback, optimized for inflow condition defined in (\ref{eq:flow_sin_base}).}\vspace{-10pt}
       \label{fig:Fig5}
\end{figure}

\begin{table}[h!]
\small
\caption{Feedback controller gains and total energy output of CCD without control load constraint for inflow condition (\ref{eq:flow_sin_base}). }\vspace{-10pt}
\label{table_3}
\begin{center}
\begin{tabular}{l c c}
\hline
\textbf{CCD Approach} & \textbf{Control Gain} & \textbf{Energy Output}
\\
 & $K_1,~K_2$ & [$J$] \\
 % & $[Nm/red/s]2$ & \\
\hline 
OLOC & N/A  & 322248 (\textbf{ref})\\\hline
Linear & 90.90 & 320312 (-0.601\%)\\
Feedback Control  & {$[Nm/(rad/s)]$} & \\\hline
Quadratic & 10.98 & 322239 (-0.003\%)\\
Feedback Control & {$[Nm/(rad/s)^2]$} & \\\hline
{Baseline Bahaj} & N/A  & 302778\\
{HKT with OLOC} &  & (-6.042\%)\\
\hline
\end{tabular}\vspace{-10pt}
\end{center}
\end{table}

Fig.~\ref{fig:Fig4} shows that the control method, i.e., OLOC or feedback, has an influence on the design space, with the quadratic feedback controller having the closest geometric design to that of CCD with OLOC. The same trend can be observed in the HKT rotational speed, control load, and fluid-induced torque, where the CCD with quadratic feedback controller results in similar trajectories as compared to the case with OLOC. The CCD with a linear feedback controller, on the other hand, shows relatively different geometric designs and performance trajectories. When comparing the energy output results in Table~\ref{table_3}, it can be seen that all three CCD cases lead to the same level of energy generation, with the linear feedback controller showing the lowest energy output, i.e., 0.6\% lower than the OLOC case. The difference between the baseline geometry and the optimized one is also noticeable. Note that the energy output of the baseline HKT \cite{bahaj2007power} with no design optimization performed with OLOC is also listed in Table~\ref{table_3}, demonstrating the benefit of CCD in increasing the energy output by 6\%.

The better performance of the quadratic feedback controller, as compared to the linear feedback controller, and its ability to accurately approximate the OLOC trajectory could be explained according to the tip-speed ratio (TSR) sub-plot in Fig.~\ref{fig:Fig5}. In the absence of a constraint on control load, it can be seen that the OLOC maintains the TSR at 7.18, which is the optimal TSR value for the CCD with OLOC case as shown in Fig.~\ref{fig:cp1}. For the CCD case with a quadratic feedback controller, the optimization attempts to find the optimal gain $K_2$ that would lead to a similar TSR as the one obtained from the OLOC. The quadratic feedback controller, constrained by its feedback structure and fixed gain, however, cannot maintain TSR at its optimal value. Such limitation in the control is even larger with the linear feedback controller, resulting in relatively large variation in the TSR as shown in Fig.~\ref{fig:Fig4}. Note that with the speed as feedback, the control trajectory has to be in phase with the rotating speed. 

Overall, the two case studies performed in this section for inflow conditions defined in (\ref{eq:simple_flow}) and (\ref{eq:flow_sin_base}) showed that, in the absence of a constraint on the control load, the OLOC trajectory can be accurately approximated by a quadratic feedback controller based on the HKT rotational speed. Such a feedback controller with the optimal gain can be directly used for real-time implementation of the closed-loop controller for the optimal geometric design. The analysis above also indicated that the effectiveness of the feedback controller structure, i.e., either linear or quadratic feedback, can be informed by CCD optimization. For the HKT application, it was shown that a quadratic feedback controller would be a more appropriate choice for the feedback controller architecture as compared to a linear feedback one.

\vspace{-3pt}
\subsection{Scenario 2: With Constraint on Control Load}

\begin{figure}[h!]
\centering
      \includegraphics[scale=0.55]{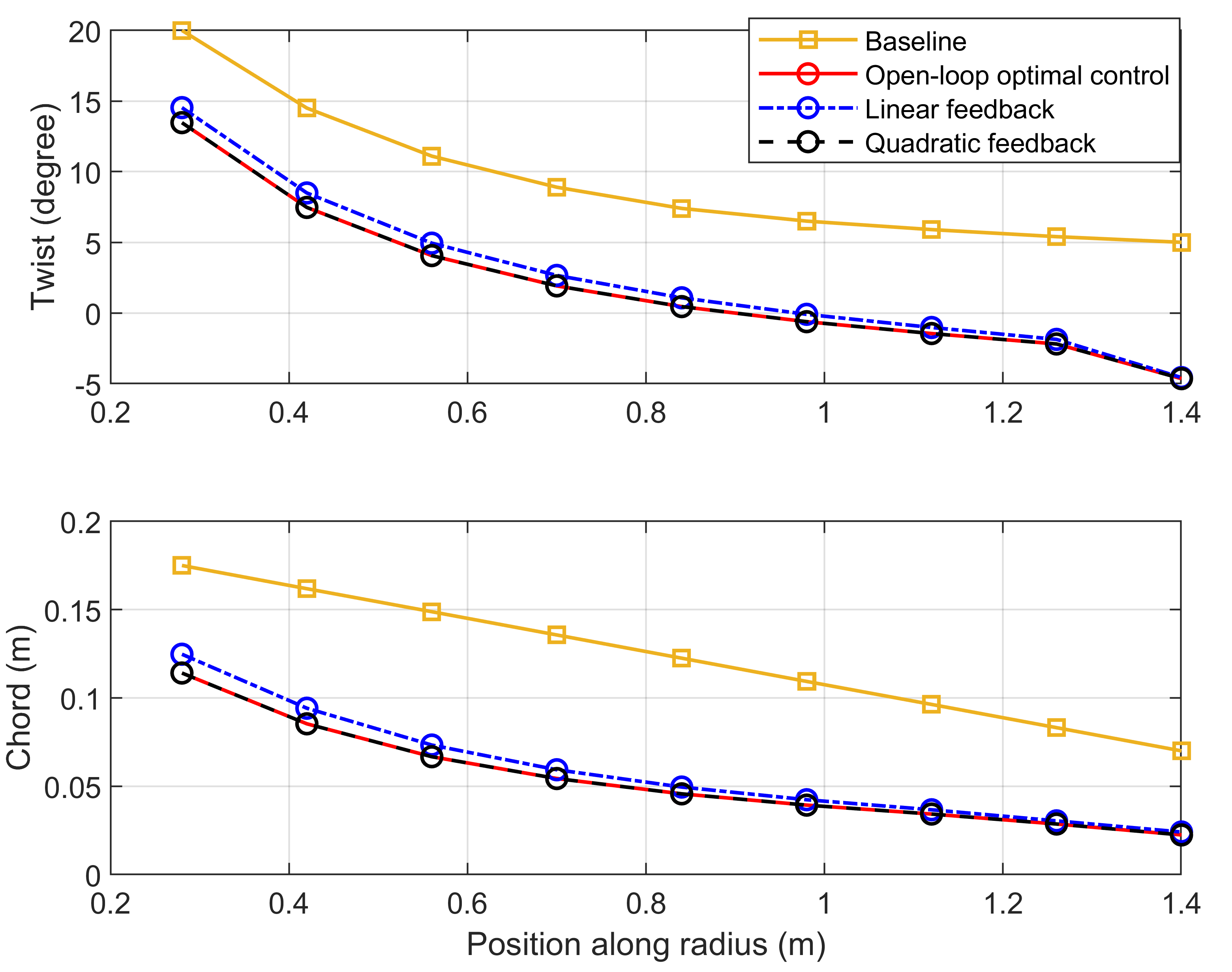}\vspace{-12pt}
      \caption{Comparison of physical geometry optimized by CCD  with OLOC, and linear and quadratic feedback controllers, as well as the baseline design {with} constraint on control for inflow condition defined in (\ref{eq:flow_sin_base}). (top) twist angle, and (bottom) chord length. }\vspace{-16pt}
       \label{fig:Fig7}
\end{figure}

In the second scenario, an upper limit on the control load of 700 $Nm$ is imposed in the CCD optimization. As discussed earlier and revealed in \cite{jiang2022control}, the addition of the control load constraint establishes a strong coupling between control and design spaces, affecting both optimal physical geometry and optimal control trajectory/gain. Additionally, in the presence of a constraint on the control load, the CCD problem may no longer be equivalent to optimal TSR tracking. For the same inflow condition defined in (\ref{eq:flow_sin_base}), the CCD optimization with OLOC and feedback controllers are repeated. The results are summarized in Figs.~\ref{fig:Fig7} and \ref{fig:Fig8} and Table~\ref{table_4}.
\vspace{-10pt}
\begin{figure}[h!]
\centering
      \includegraphics[scale=0.5]{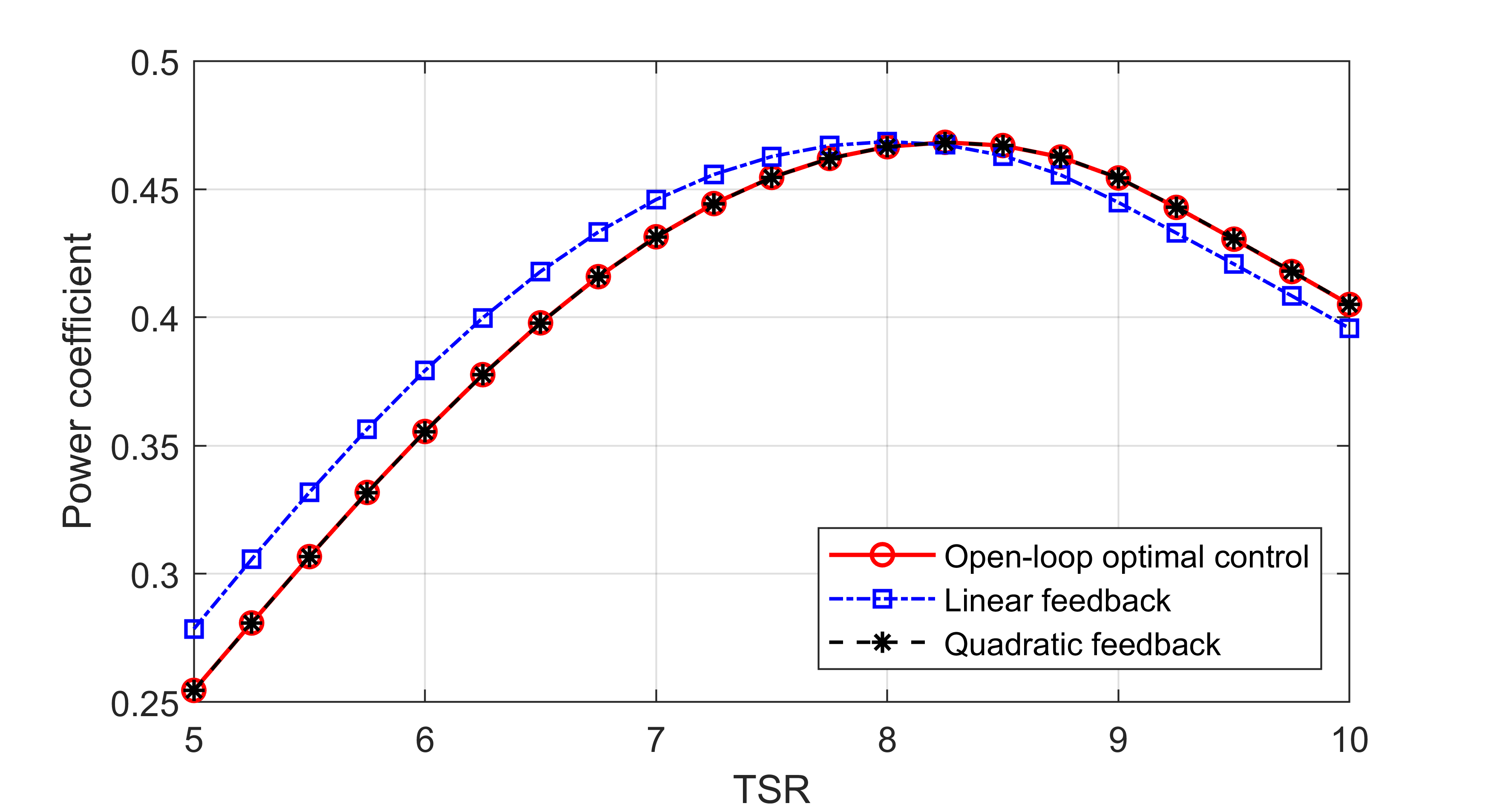}\vspace{-10pt}
      \caption{{Power coefficients versus TSR for CCD optimization with OLOC, linear and quadratic feedback with constraint on control load.}}\vspace{-10pt}
       \label{fig:cp2}
\end{figure}

\vspace{-6pt}
\begin{figure}[h!]
\centering
      \includegraphics[scale=0.6]{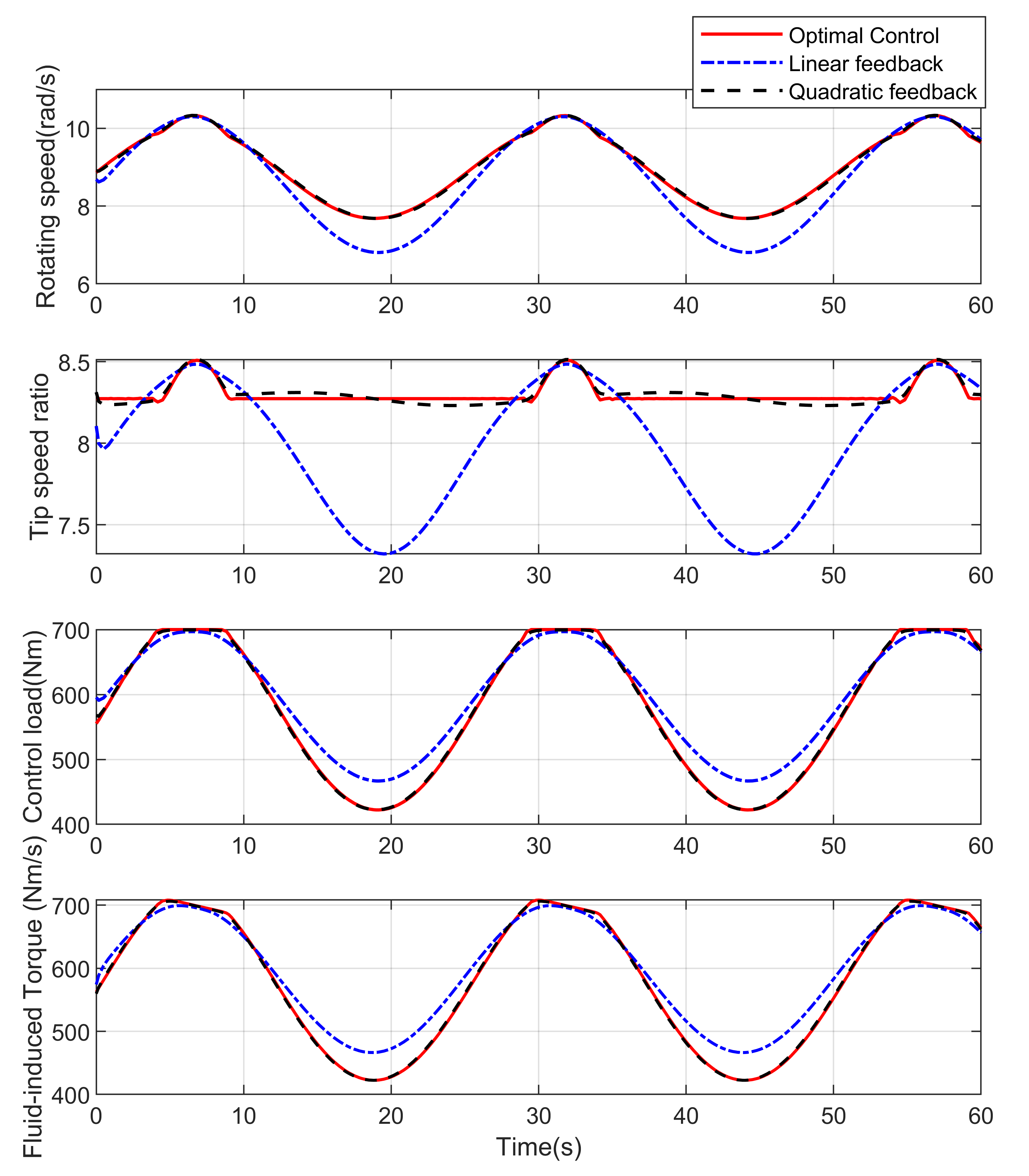}\vspace{-5pt}
      \caption{HKT rotating speed, tip speed ratio, control load, and fluid-induced torque resulted from CCD optimization with OLOC, linear feedback, and quadratic feedback, optimized for inflow condition defined in (\ref{eq:flow_sin_base}) {with} constraint on control.}\vspace{-10pt}
       \label{fig:Fig8}
\end{figure}
\begin{table}[h!]
\small
\caption{Feedback controller gains and total energy output of CCD with control load constraint for inflow condition (\ref{eq:flow_sin_base}).}\vspace{-10pt}
\label{table_4}
\begin{center}
\begin{tabular}{l c c}
\hline
\textbf{CCD Approach} & \textbf{Control Gain} & \textbf{Energy Output}
\\
 & $K_1,~K_2$ & [$J$] \\
 % & \yled{$[Nm/rad/s]2$} & \\
\hline
OLOC & N/A  & 321316 (\textbf{ref})\\\hline
Linear & 68.64  & 319443 (-0.6\%)\\
Feedback Control& {$[Nm/(rad/s)]$} & \\\hline
Quadratic & 7.16 & 321294 (-0.007\%)\\
Feedback Control& {$[Nm/(rad/s)^2]$} & \\
\hline
\end{tabular}\vspace{-10pt}
\end{center}
\end{table}

Comparing Fig.~\ref{fig:Fig8} and Fig.~\ref{fig:Fig5} also shows that the constraint on the control load affects the rotational speed, TSR, control load, and fluid-induced torque. As for the CCD with OLOC, the optimal TSR  shown in Fig. \ref{fig:cp2}, i.e., {8.27}, cannot be maintained at all times due to the hard limit on the control load. Surprisingly, the CCD with the quadratic feedback controller is still able to approximate the OLOC trajectory even with a fixed feedback gain. This enables the quadratic feedback controller to generate energy and perform very closely to the CCD case with OLOC (only 0.007\% lower energy output). Similar to Sec.~\ref{sec:sce1} results, the CCD with a linear feedback controller has lower energy output (0.7\% lower than OLOC) with a relatively distinctive geometric design profile and performance trajectories. Since for the linear feedback controller the rotational speed is explicitly used, the rotational speed, control load, and the fluid-induced torque have sinusoidal profiles, consistent with the sinusoidal profile of the inflow speed. The quadratic feedback controller, on the other hand, uses nonlinear feedback ($\omega^2$), providing it with the possibility to apply non-sinusoidal control loads, which helps to reach results much closer to the OLOC case.

In summary, the main conclusions reached in Sec.~\ref{sec:OLPC} are consistent with those observed in this section with the constraint on the control load. The CCD results confirm that the OLOC trajectory can be approximated accurately using a quadratic feedback controller, resulting in having a feedback controller with an optimized feedback gain that generates the same level of energy while having a much simpler structure and being a more suitable choice for real-time implementation in closed-loop for the final optimized HKT design, as compared to the linear feedback controller. Moreover, one natural benefit of a feedback controller for real-time implementation is its inherent robustness to disturbances and uncertainties. To investigate such a benefit, in the next section, the HKT physical design obtained from the CCD optimization is simulated with feedback controllers under uncertainties in flow conditions.

\section{Sensitivity Analysis to Uncertainties in Flow Conditions}
\label{sec:sec_5}
In this section, the HKT physical designs obtained from CCD optimizations with three different controllers are used as the actual turbine plant model, and the corresponding controller is simulated along the plant model. For the OLOC, the control trajectory rechorded from CCD optimization is applied to the HKT. For the other two designs, the HKT plant model is put in a closed-loop with a feedback controller that reads the rotational speed of the HKT as the feedback signal.
The control input to the HKT plant model is determined using either linear or quadratic laws defined in (\ref{eq:FB_controller}) and (\ref{eq:FB_controller_s}), respectively.

It is assumed that the turbine speed sensor has a Gaussian noise, which only influences the feedback controllers. The actual feedback signal ($\Bar{\omega}$) that is being fed to the feedback controllers is calculated as {$\bar{\omega} = F[\omega + n]$}, where $n$ is a {pre-assumed} zero-mean Gaussian noise, {which has an averaged signal-to-noise ratio (SNR) of 20dB compared with the turbine rotational speed signal}, and $F$ is a 2nd-order Butterworth low-pass filter with cutoff frequency of $0.5~Hz$. In addition to rotational speed sensor noise, we define three forms of uncertainties in flow conditions:
\begin{itemize}
    \item \textbf{Type A}: 10\% uncertainty on inflow amplitude, \\$v_A(t)=1.1v_0=1.1(0.2 \sin(0.25t) + 1.5)$,
    \item \textbf{Type B}: uncertainty on inflow frequency, \\$v_B(t)=0.2 \sin(0.225t) + 1.5$, 
    \item \textbf{Type C}: uncertainty on inflow phase, \\$v_C(t)=0.2 \sin(0.25t+0.2\pi) + 1.5$. 
\end{itemize}

Note that $v_0$ from (\ref{eq:flow_sin_base}) was used during the CCD process. However, when the HKT design is evaluated, it is exposed to {different inflow conditions} defined by Types A, B, and C above. The feedback controller is able to react to the actual inflow through the feedback mechanism, but the OLOC is the same as the one obtained from CCD. For bench-marking purpose, an additional case is considered where an OLOC with the exact knowledge of the inflow condition is solved and applied to HKT. This case would provide the ceiling on the maximum energy output that can be harvested for different flow conditions.

\begin{figure}[h!]
\centering
      \includegraphics[scale=0.6]{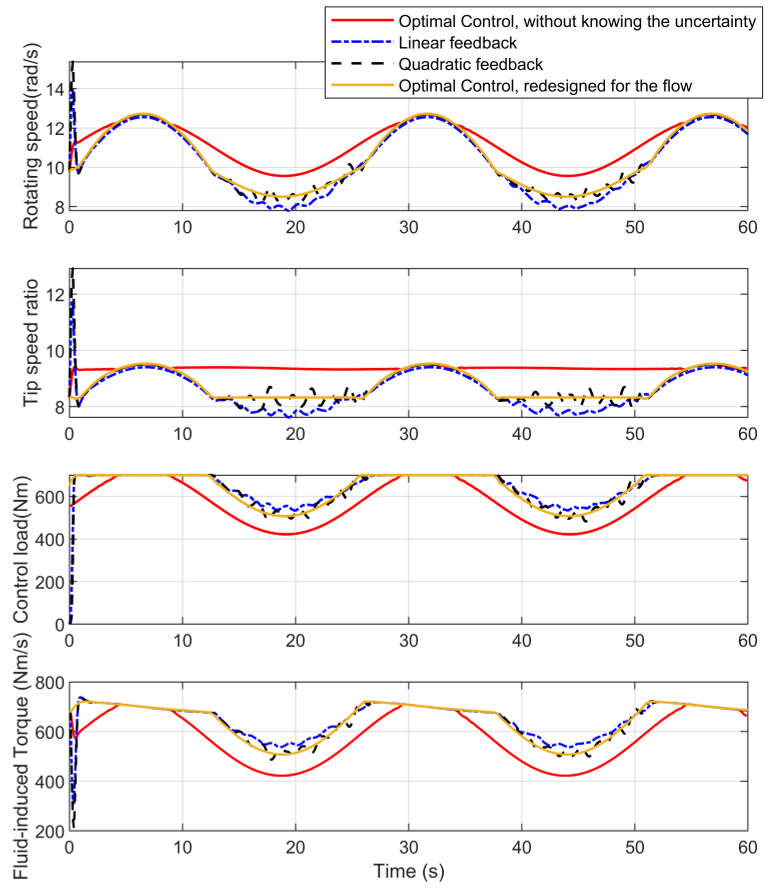}\vspace{-12pt}
      \caption{HKT performance resulted from CCD optimization with OLOC, linear feedback, and quadratic feedback {with} constraint on control and {with} Type A on flow condition.}\vspace{-16pt}
       \label{fig:Fig10}
\end{figure}

\begin{figure}[h!]
\centering
      \includegraphics[scale=0.6]{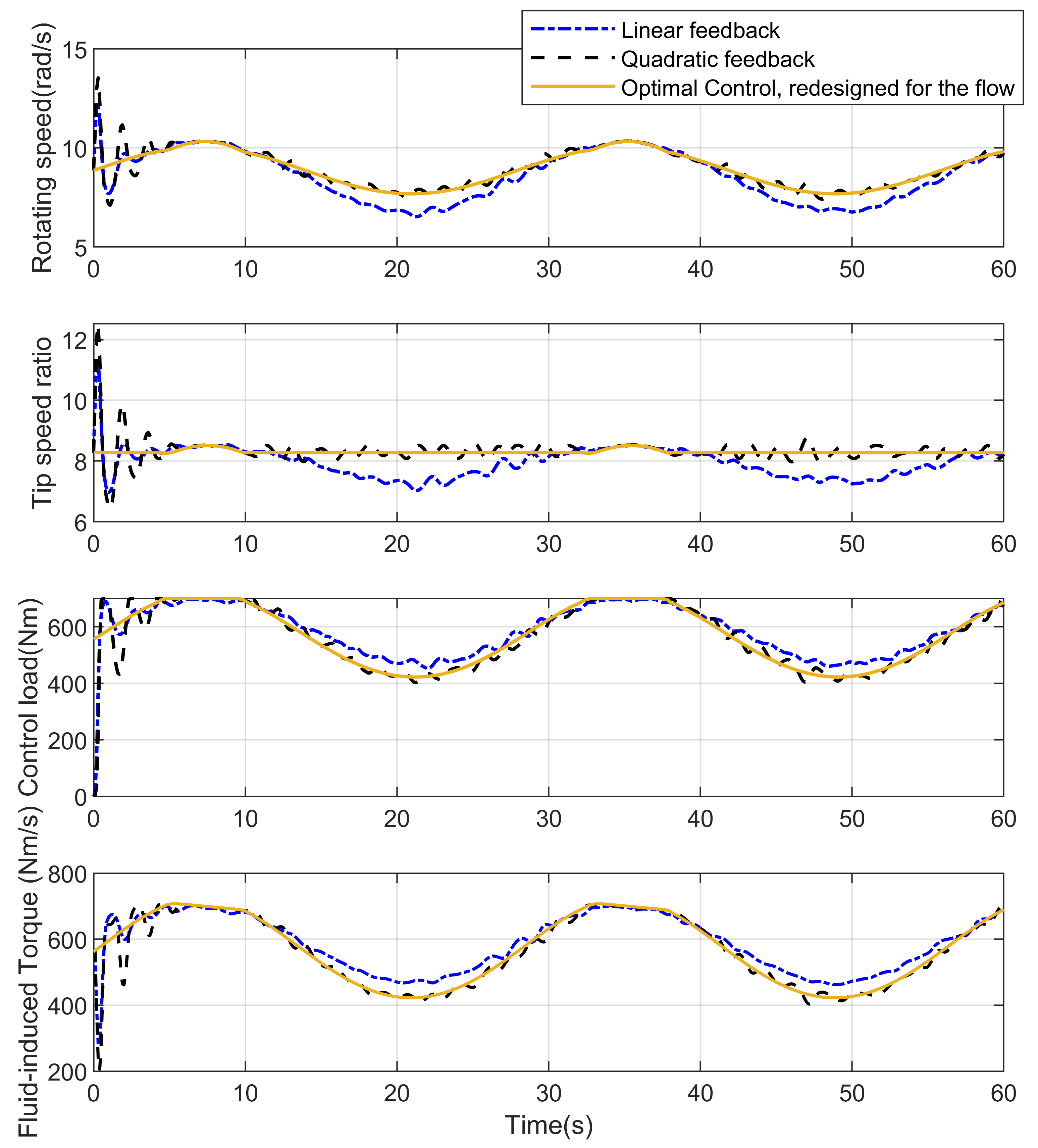}\vspace{-12pt}
      \caption{HKT performance resulted from CCD optimization with OLOC, linear feedback, and quadratic feedback {with} a constraint on control and {with} Type B {on flow} condition.}\vspace{-14pt}
       \label{fig:Fig11}
\end{figure}

\begin{figure}[h!]
\centering
      \includegraphics[scale=0.6]{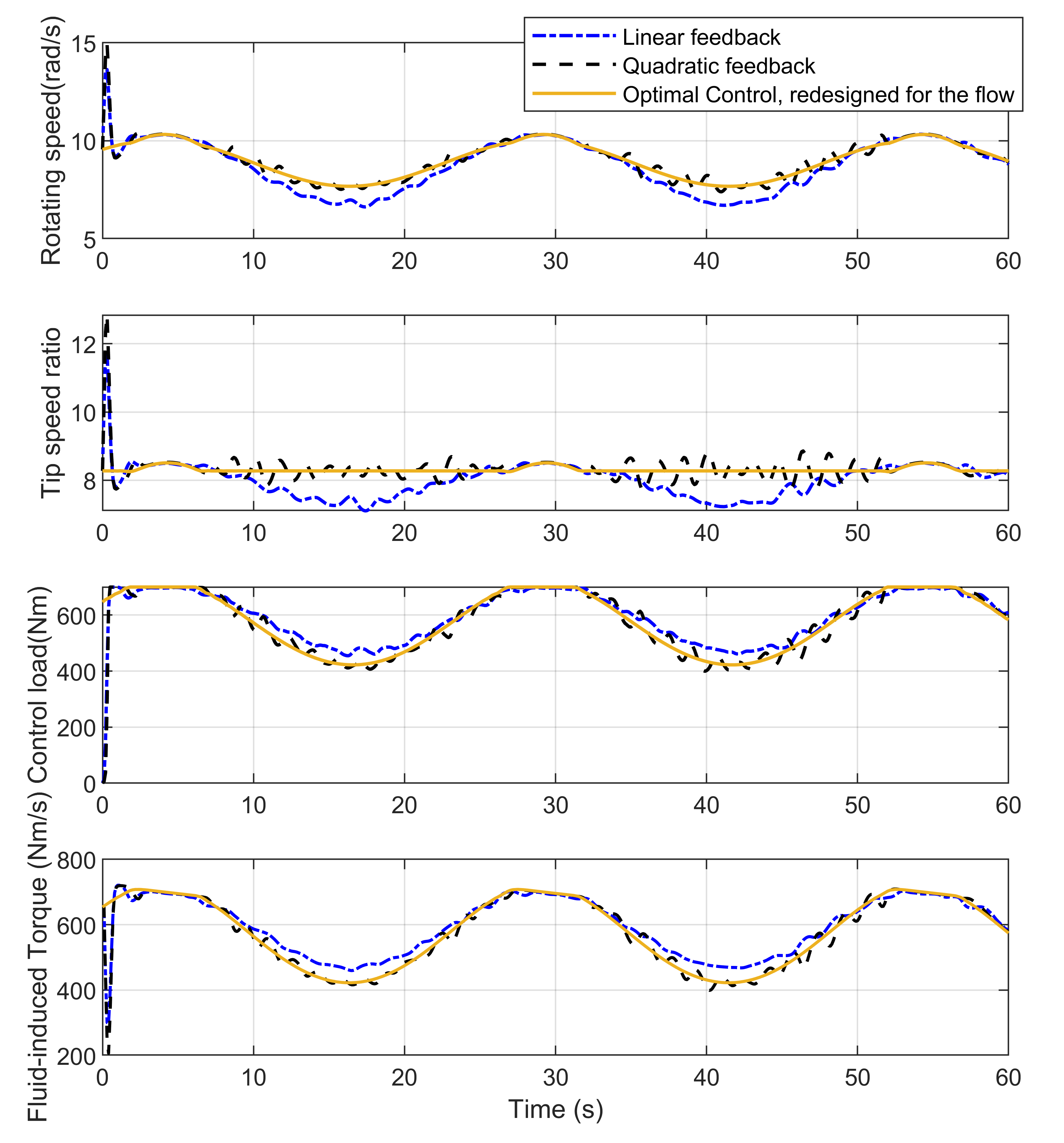}\vspace{-12pt}
      \caption{HKT performance resulted from CCD optimization with OLOC, linear feedback, and quadratic feedback {with} constraint on control and {with} uncertainty Type C on flow condition.}\vspace{-7pt}
       \label{fig:Fig12}
\end{figure}

The rotational speed, constrained control load, and fluid-induced torque of the designs obtained from the CCD are simulated with different controllers.
The results are plotted in Figs.~\ref{fig:Fig10}, \ref{fig:Fig11}, and \ref{fig:Fig12} for flow uncertainty Types A, B, and C, respectively. The energy output results are summarized for all cases in Table~\ref{table_5}. Note that in Figs.~\ref{fig:Fig11} and \ref{fig:Fig12}, the {original} OLOC case is not shown as the HKT simulation with the {original} OLOC trajectory under Types B and C of flow condition uncertainty leads to {a negative rotational speed}, which is not a feasible scenario and may mean the turbine operation is unstable with no energy generation.
%
%\vspace{-6pt}
%
%\vspace{-6pt}
%\vspace{-6pt}

\begin{table}[h!]
\small
\caption{Feedback controller gains and total energy output from the final HKT design reached by CCD with OLOC, linear and quadratic feedback controllers {with} constraint on control and {with} uncertainty Types A, B, and C in flow condition. }\vspace{-10pt}
\label{table_5}
\begin{center}
\begin{tabular}{l c c c}
\hline
\textbf{Control} & & \textbf{Energy Output} & 
\\
\textbf{Approach} & & $[J]$ & 
\\
 & Type A & Type B & Type C\\
\hline
OLOC & 398973  & N/A$^{*}$ & N/A$^{*}$\\
 & (-3.7\%)  &  & \\\hline
Linear & {409807} & {307622}  & {319226}  \\
Feedback Control &  ({-1.0 \%}) & ({-0.9 \%}) & ({-0.8 \%})\\\hline
Quadratic & {412436}  & {308495}  & {320570}  \\
Feedback Control &  ({-0.5 \%}) & ({-0.5 \%}) & ({-0.4 \%})\\\hline
OLOC  & 414301 & 310307 & 321936\\
Redesigned & (\textbf{ref}) & (\textbf{ref}) & (\textbf{ref})\\
%uncertainty & (\textbf{ref}) & (\textbf{ref}) & (\textbf{ref})\\
\hline
\multicolumn{4}{l}{\scriptsize * HKT rotational velocity drops down to negative speed, not generating energy.}\vspace{-2pt} \\
%\multicolumn{4}{l}{\scriptsize ~~turbine does not generate energy.} \\
\end{tabular}\vspace{-10pt}
\end{center}
\end{table}

For the flow uncertainty Type A, the OLOC has the worst energy output as listed in Table~\ref{table_5}. This is not surprising as the OLOC is sensitive to {deviation from design conditions}. Both feedback controllers, on the other hand, show performance trajectories close to those obtained from OLOC with {the knowledge of inflow condition deviations}. Specifically, it is observed from Figs.~\ref{fig:Fig10}, \ref{fig:Fig11}, and \ref{fig:Fig12} that the quadratic feedback controller, as compared to the linear feedback controller, demonstrates much closer performance to the OLOC with {knowledge of inflow condition deviations}. In terms of energy output, Table~\ref{table_5} shows that the linear feedback controller is able to reach 99-99.2\% of the energy generation ceiling, while the quadratic controller can reach 99.5-99.6\% of the energy generation ceiling.

Overall, the sensitivity analysis to uncertainties in flow conditions confirmed (i) the robustness of the feedback controller against such uncertainties and (ii) the superior performance of the quadratic controller as compared to the linear feedback controller. The latter conclusion was previously reached during the CCD process, showing that the CCD can be also leveraged to determine the best feedback control architecture with known structures by comparing the OLOC with different feedback control architectures.

\vspace{-0.15cm}   
\section{Summary and Conclusions}
\label{sec:sec_6}

In the paper, we expanded our previous development to enable control co-design (CCD) of hydrokinetic turbines (HKT) with a feedback controller. With the new development, we conducted a comparison of hydrokinetic turbine CCD with open-loop optimal control (OLOC) and feedback controllers. Two feedback control laws were investigated: one considers a proportional linear relationship between the generator control load and the turbine rotational speed, while the other assumes a square relationship, which is also known as the ``$K\omega^2$'' law.

{The comparison includes two sets of scenarios: without generator control load constraint and with control load constraint.
Consistent with our previous study, introducing control load constraints change the optimal physical design variables.
More importantly, results from both scenarios show that different feedback controllers lead to different physical designs. Among the two investigated feedback control laws, the $K\omega^2$ control law better resembles the performance of the OLOC case, compared to the linear case, in achieving optimal tip speed ratio tracking.} 

{To demonstrate the robustness of the feedback controller designed through the CCD process, we further conducted sensitivity analyses to inflow condition variations for the final HKT design simulated with a feedback controller. These sensitivity analyses confirmed (i) the robustness of the feedback
controller against inflow uncertainties and (ii) the superior performance of the quadratic feedback controller as compared to the linear feedback controller.}

\section*{Acknowledgment}
This work is supported by the United States
Department of Energy (DOE)-ARPA-E under SHARKS program award No. DE-AR0001438.

\vspace{-0.2cm}
\bibliographystyle{unsrt} % unsrt
\bibliography{ACC2023Ref.bib}

\begin{thebibliography}{10}

\bibitem{fathy2001coupling}
H.K. Fathy, J.A. Reyer, P.Y. Papalambros, and A.G. Ulsov.
\newblock On the coupling between the plant and controller optimization
  problems.
\newblock In {\em American Control Conference (ACC)}, volume~3, pages
  1864--1869, 2001.

\bibitem{peters2009measures}
D.L. Peters, P.Y. Papalambros, and A.G. Ulsoy.
\newblock On measures of coupling between the artifact and controller optimal
  design problems.
\newblock In {\em International Design Engineering Technical Conferences and
  Computers and Information in Engineering Conference}, volume 48999, pages
  1363--1372, 2009.

\bibitem{patil2012computationally}
R.~Patil, Z.~Filipi, and H.~Fathy.
\newblock Computationally efficient combined plant design and controller
  optimization using a coupling measure.
\newblock {\em Journal of Mechanical Design}, 134(7):071008, 2012.

\bibitem{garcia2019control}
M.~Garcia-Sanz.
\newblock Control co-design: an engineering game changer.
\newblock {\em Advanced Control for Applications: Engineering and Industrial
  Systems}, 1(1):e18, 2019.

\bibitem{ross2022development}
H.~Ross, M.~Hall, D.R. Herber, J.~Jonkman, A.K. Sundarrajan, T.T. Tran,
  A.~Wright, D.~Zalkind, and N.~Johnson.
\newblock Development of a control co-design modeling tool for marine
  hydrokinetic turbines.
\newblock Technical report, National Renewable Energy Lab.(NREL), Golden, CO
  (United States), 2022.

\bibitem{kimball2022results}
R.~Kimball, A.~Robertson, M.~Fowler, N.~Mendoza, A.~Wright, A.~Goupee,
  E.~Lenfest, and A.~Parker.
\newblock Results from the focal experiment campaign 1: turbine control
  co-design.
\newblock In {\em Journal of Physics: Conference Series}, volume 2265, page
  022082. IOP Publishing, 2022.

\bibitem{deese2018nested}
J.~Deese and C.~Vermillion.
\newblock Nested plant/controller codesign using g-optimal design and
  continuous time adaptation laws: Theoretical framework and application to an
  airborne wind energy system.
\newblock {\em Journal of Dynamic Systems, Measurement, and Control}, 140(12),
  2018.

\bibitem{deodhar2016framework}
N.~Deodhar and C.~Vermillion.
\newblock A framework for fused experimental/numerical plant and control system
  optimization using iterative g-optimal design of experiments.
\newblock In {\em International Design Engineering Technical Conferences and
  Computers and Information in Engineering Conference}, volume 50107, page
  V02AT03A010. American Society of Mechanical Engineers, 2016.

\bibitem{baheri2017combined}
A.~Baheri, J.~Deese, and C.~Vermillion.
\newblock Combined plant and controller design using bayesian optimization: A
  case study in airborne wind energy systems.
\newblock In {\em Dynamic Systems and Control Conference}, volume 58295, page
  V003T40A003. American Society of Mechanical Engineers, 2017.

\bibitem{naik2021fused}
K.~Naik, S.~Beknalkar, A.~Mazzoleni, and C.~Vermillion.
\newblock Fused geometric, structural, and control co-design framework for an
  energy-harvesting ocean kite.
\newblock In {\em 2021 American Control Conference (ACC)}, pages 3525--3531,
  2021.

\bibitem{jiang2022control}
B.~Jiang, M.R. Amini, Y.~Liao, J.R.R.A. Martins, and J.~Sun.
\newblock {Control Co-design of a Hydrokinetic Turbine with Open-loop Optimal
  Control}.
\newblock {\em ASME 41st International Conference on Ocean, Offshore and Arctic
  Engineering (OMAE2022)}, 2022.
\newblock {Hamburg, Germany}.

\bibitem{pao2021control}
L.Y. Pao, D.S. Zalkind, D.T. Griffith, M.~Chetan, M.S. Selig, G.K. Ananda, C.J.
  Bay, T.~Stehly, and E.~Loth.
\newblock Control co-design of 13 mw downwind two-bladed rotors to achieve 25\%
  reduction in levelized cost of wind energy.
\newblock {\em Annual Reviews in Control}, 2021.

\bibitem{deshmukh2013simultaneous}
A.~Deshmukh and J.~Allison.
\newblock Simultaneous structural and control system design for horizontal axis
  wind turbines.
\newblock In {\em 54th AIAA/ASME/ASCE/AHS/ASC structures, structural dynamics,
  and materials conference}, page 1533, 2013.

\bibitem{naik2022fused}
K.P. Naik.
\newblock {\em Fused Site, Plant, and Control System Co-Design with Application
  to Underwater Energy-Harvesting Kites}.
\newblock PhD thesis, North Carolina State University, 2022.

\bibitem{sundarrajan2021open}
A.K. Sundarrajan, Y.H. Lee, J.T. Allison, and D.R. Herber.
\newblock Open-loop control co-design of floating offshore wind turbines using
  linear parameter-varying models.
\newblock In {\em International Design Engineering Technical Conferences and
  Computers and Information in Engineering Conference}, page V03AT03A010.
  American Society of Mechanical Engineers, 2021.

\bibitem{du2020control}
X.~Du, L.~Burlion, and O.~Bilgen.
\newblock Control co-design for rotor blades of floating offshore wind
  turbines.
\newblock In {\em ASME International Mechanical Engineering Congress and
  Exposition}, page V07AT07A052, 2020.

\bibitem{coe2020initial}
R.G. Coe, G.~Bacelli, S.~Olson, V.S. Neary, and M.~Topper.
\newblock Initial conceptual demonstration of control co-design for wec
  optimization.
\newblock {\em Journal of Ocean Engineering and Marine Energy}, 6(4):441--449,
  2020.

\bibitem{naik2022combined}
K.~Naik, S.~Beknalkar, J.~Reed, A.~Mazzoleni, H.~Fathy, and C.~Vermillion.
\newblock Combined plant and controller optimization of an underwater energy
  harvesting kite system.
\newblock {\em arXiv preprint arXiv:2206.08472}, 2022.

\bibitem{deshmukh2015bridging}
A.P. Deshmukh, D.R. Herber, and J.T. Allison.
\newblock Bridging the gap between open-loop and closed-loop control in
  co-design: A framework for complete optimal plant and control architecture
  design.
\newblock In {\em American Control Conference (ACC)}, pages 4916--4922, 2015.

\bibitem{nash2021robust}
A.L. Nash, H.C. Pangborn, and N.~Jain.
\newblock Robust control co-design with receding-horizon mpc.
\newblock In {\em 2021 American Control Conference (ACC)}, pages 373--379,
  2021.

\bibitem{bahaj2007power}
A.S. Bahaj, A.F. Molland, J.R. Chaplin, and W.~Batten.
\newblock Power and thrust measurements of marine current turbines under
  various hydrodynamic flow conditions in a cavitation tunnel and a towing
  tank.
\newblock {\em Renewable Energy}, 32(3):407--426, 2007.

\bibitem{ning2014simple}
S.A. Ning.
\newblock A simple solution method for the blade element momentum equations
  with guaranteed convergence.
\newblock {\em Wind Energy}, 17(9):1327--1345, 2014.

\bibitem{garcia2021engineering}
M.~Garcia-Sanz.
\newblock Engineering microgrids with control co-design: Principles, methods,
  and metrics.
\newblock {\em IEEE Electrification Magazine}, 9(3):8--17, 2021.

\bibitem{kim2012maximum}
K-H. Kim, T.L. Van, D-C. Lee, S-H. Song, and E-H. Kim.
\newblock Maximum output power tracking control in variable-speed wind turbine
  systems considering rotor inertial power.
\newblock {\em IEEE Transactions on Industrial Electronics}, 60(8):3207--3217,
  2012.

\bibitem{kim2010fast}
K-H. Kim, D-C. Lee, and J-M. Kim.
\newblock Fast tracking control for maximum output power in wind turbine
  systems.
\newblock In {\em 20th Australasian Universities Power Engineering Conference},
  pages 1--5, 2010.

\bibitem{falck2021dymos}
R.~Falck, J.S. Gray, K.~Ponnapalli, and T.~Wright.
\newblock Dymos: A python package for optimal control of multidisciplinary
  systems.
\newblock {\em Journal of Open Source Software}, 6(59):2809, 2021.

\bibitem{Abbas2022}
J.A. Nikhar, S.Z. Daniel, P.~Lucy, and W.~Alan.
\newblock A reference open-source controller for fixed and floating offshore
  wind turbines.
\newblock {\em Wind Energy Science}, 7(1):53--73, January 2022.

\bibitem{Bossanyi2000}
E.~A. Bossanyi.
\newblock The design of closed loop controllers for wind turbines.
\newblock {\em Wind Energy}, 3(3):149--163, July 2000.

\bibitem{Johnson2006}
K.E. Johnson, L.Y. Pao, M.J. Balas, and L.J. Fingersh.
\newblock Control of variable-speed wind turbines: standard and adaptive
  techniques for maximizing energy capture.
\newblock {\em IEEE Control Systems Magazine}, 26(3):70--81, 2006.

\bibitem{avvakumov2000boundary}
S.N. Avvakumov and Y.~Kiselev.
\newblock Boundary value problem for ordinary differential equations with
  applications to optimal control.
\newblock In {\em World Multi-Conf. on Systemics Cybernetics and Informatics},
  2004.

\bibitem{gill2005snopt}
P.E. Gill, W.~Murray, and M.A. Saunders.
\newblock Snopt: An sqp algorithm for large-scale constrained optimization.
\newblock {\em SIAM Review}, 47(1):99--131, 2005.

\bibitem{ning2013ccblade}
S.A. Ning.
\newblock {\em CCBlade documentation}.
\newblock National Renewable Energy Laboratory, 2013.

\end{thebibliography}

\end{document}